\documentclass[aps,prx,reprint,amsmath,amssymb,nofootinbib]{revtex4-2}
\usepackage{graphicx}
\usepackage{bm}

\begin{document}

\title{The thermodynamic budget of objectivity}

\author{Wonyong Chung}
\affiliation{Department of Physics, Princeton University, Princeton, New Jersey 08544, USA}

\date{\today}

\begin{abstract}
An event becomes a fact when enough independent records of it exist such that no operation is possible that leaves all observers ignorant of it again. Records are physical, and creating, keeping, and destroying them have thermodynamic costs. We compute these costs exactly in a minimal broadcast model with a driven central qubit writing
records into $N$ initially thermal registers, with work sources and baths at temperature
$T$. The basic arguments for the operations are outlined here, with full treatment in the text. Creation of one event record at redundancy $R$ and information deficit $\delta$
requires work $W \ge R\, k_\mathrm{B}T \ln 2\, (1-\delta) H(\varepsilon)$, linear in
redundancy at the Landauer coefficient. Recording in a basis misaligned by
$\theta$ with the einselecting interaction fails at a critical angle,
$\theta_c = 2\arcsin\sqrt{H_2^{-1}(\delta)}$, beyond which objectivity is
unreachable at any work, and below $\theta_c$ the optimal work is a closed form in the pointer basis that minimizes the work of objectivity. Reducing objectivity is impossible at any work with access to fewer than $R - R_{\min} + 1$ records, while work protects nothing if given access. Inter-record redundancy is extractable free energy and erasing the accessed records yields a net gain of work beyond the first copy.
The second law constrains the destruction of the correlation: erasing a fact from a sector an adversary controls dissipates at least the mutual information erased, but a consolidation dissipates nothing.
Records in re-thermalizing media are nonequilibrium steady states, and the
minimal power to maintain a record at $(R,\delta)$ is the known two-state driving bound evaluated
at the record-quality boundary.
On the propagation of records: no unitary operation creates a record in a zero-gap thermal register (i.e. every record is written through the bath) and independently written records obey
$\Sigma A \ge R^2(1-2e^*)^2/2$, with $\Sigma$ the entropy production and $A$ the
dynamical activity. The average
rate at which classicality can emerge is set by the geometric mean of dissipation and
bath activity. We concretize the framework in the context of two instrument classes: a total-absorption
calorimeter with a per-event information budget, and an adiabatic magnetic drift
filter in which an electron radiates
${\approx}0.66\,\mathrm{meV}$ as ${\sim}9$ microwave photons that outrun the emitting electron by orders of magnitude. Detectors, in this framework, are devices that meet the costs of objective measurement by
design and we propose the thermodynamic budget of objectivity as an instrument design quantity.
\end{abstract}

\maketitle

\section{Introduction}\label{sec:intro}

Classical properties are objective: many observers can find them out independently,
agree, and leave the system intact. Quantum Darwinism locates the mechanism in
redundancy---the environment that decoheres a system also monitors its pointer
observables~\cite{Zurek1981,Zurek1982} and broadcasts many records of them, so that
observers read fragments of the environment rather than the system
itself~\cite{Zurek2009,OllivierPoulinZurek2004,OllivierPoulinZurek2005,BlumeKohoutZurek2005,BlumeKohoutZurek2006,BlumeKohoutZurek2008}.
An event that``happened, factually'' is then a technical statement of operations: recorded, redundantly, in
fragments many observers can independently reach. But records are physical, and
information processing has thermodynamic costs~\cite{Landauer1961}. Creating a record,
holding it against a thermalizing environment, and destroying it are physical
processes, and the quantum-thermodynamics
roadmap's perspective on the thermodynamics of information propagation poses, as an open
problem, ``is there a thermodynamic price associated with the
emergence of classicality?''~\cite{Roadmap2026}.

Debarba, Huber, and
Friis~\cite{DebarbaHuberFriis2024} bound the entropy production of redundant copying
from below by the mixedness of the memories written into---the Holevo--Landauer bound,
saturated on spectrum-broadcast states among other commuting record structures.
Candeloro, Debarba, and
Binder~\cite{CandeloroDebarbaBinder2026} constrain intersubjective agreement through the
third law. Schwarzhans et~al.~\cite{Schwarzhans2026} build detectors as autonomous
thermal machines, with detection quality tied to entropy production. The
composition of the budget itself has not yet been stated: a law for the work of creating objectivity, with
attainability; a speed limit for its emergence; an adversary analysis at bounded access
and bounded work; the growth of the access requirement as an event ages; the
maintenance cost of objectivity per unit time; and per-event energetics for a real
instrument. We compute this composition in a minimal model and make
it concrete in two working instruments whose per-event information budgets are
computed.

The composition is built from structures the two parent fields already hold. Erasing a
bit costs at least $k_\mathrm{B}T\ln 2$~\cite{Landauer1961}, measurement can defer its
cost to apparatus reset~\cite{SagawaUeda2008,SagawaUeda2009}, and the nonequilibrium
accounting carries the final state's free energy above
thermal~\cite{EspositoVandenBroeck2011,ReebWolf2014}. Creating correlations against
thermal surroundings costs work~\cite{Huber2015}; conversely, correlations store
extractable work~\cite{PerarnauLlobet2015,delRio2011,Bera2017}. When a system exchanges
information with an untouched partner, the partner-inclusive entropy production carries
the change in mutual information---the second law with information
exchange~\cite{SagawaUeda2012,ParrondoHorowitzSagawa2015}---and holding a state against
its own relaxation requires a known minimum power~\cite{HorowitzZhouEngland2017}. On the
Darwinism side, only pointer observables are redundantly
recorded~\cite{OllivierPoulinZurek2004,OllivierPoulinZurek2005,ZwolakZurek2013}, records
suppress conjugate coherence exponentially in their number~\cite{Zurek1982}, and
redundancy emerges generically in decohering
dynamics~\cite{BrandaoPianiHorodecki2015,Korbicz2014}. Genericity, though, rides on free
energy already flowing---illumination, scattering; the question of cost asks what is
\emph{minimal}, not what is typical.

Why pose the question now, and why from instrumentation? Because the pieces have
independently matured: measurement thermodynamics has exact fluctuation relations and
experiments~\cite{SagawaUeda2010,Toyabe2010}; Darwinism has moved from interpretation to
theorems~\cite{BrandaoPianiHorodecki2015,Korbicz2014,Korbicz2021} and to
experiments~\cite{Ciampini2018,Chen2019,Unden2019}; and detector physics has begun
computing complete per-event information budgets for real
instruments~\cite{ChungCalorimetry2026,ChungQuantumFilter2026}, which is precisely the quantity the
question needs on its cost side. When a calorimeter's record of a shower and
an electromagnetic filter's record of an electron are quantified channel by channel, ``what did it cost
for this event to become known'' becomes a question that has a definitive answer.

\section{Records, objectivity, and costs}\label{sec:definitions}

Let $S$ be the system carrying the event: a state change $\varepsilon$ (minimally, one
bit) at time $t_0$. Let the environment be partitioned into fragments
$E_1,\ldots,E_N$.

\begin{itemize}
\item \textbf{Record.} Fragment $E_i$ holds a record of $\varepsilon$ at deficit
$\delta$ if the accessible mutual information between $\varepsilon$ and $E_i$ is at
least $(1-\delta)$ of $\varepsilon$'s entropy. A record is what the fragment
\emph{retains} about the event, never the interaction that wrote it---the same
retained-description convention as the calorimetric information
budget~\cite{ChungCalorimetry2026}. The state-level characterization of \emph{exact} shared
records (nondisturbing, agreement with probability one) is Gu's spectrum-sharing
structure~\cite{Gu2026}; the records whose costs are computed here are the
information-level, deficit-graded object.

\item \textbf{Objectivity at $\bm{(R,\delta)}$.} At least $R$ disjoint fragments each
hold a record at deficit $\delta$---the Darwinism redundancy $R_\delta$, used
operationally: $R$ independent observers can each learn the fact without touching $S$
or each other. For $R$ conditionally independent symmetric records at flip error $e$,
the deficit converts exactly to the agreement and bias metrics of thermodynamic
intersubjectivity~\cite{CandeloroDebarbaBinder2026}:
$\mathrm{Agr} = (1-e)^R + e^R$ and $\mathrm{Bias} = e\,|1-2p|$, with $p$ the event's
prior---an invertible map within this family ($R \ge 2$, $0 \le e \le 1/2$,
nondegenerate fixed prior). Across record families the two coordinates are not
interconvertible: agreement can be perfect while event information is zero, and
conversely; the one-way bound, from shared-record success to accessible-information
consensus, is Gu's~\cite{Gu2026}.

\item \textbf{Recording process.} Any physical process that takes the environment from
its prior state to one with objectivity at $(R,\delta)$, using a work source and baths
at temperature $T$.

\item \textbf{Un-knowing as two inequivalent notions.} An adversary class
$\mathcal{U}(k,W)$ has access to at most $k$ fragments plus its own ancillas and memory
(all counted), work at most $W$, and free thermal contact. \emph{Redundancy
reduction} succeeds when objectivity falls below a stated threshold: no
$R' \ge R_{\min}$ fragments remain at deficit $\delta'$ accessible to future observers.
\emph{Erasure from the sector} succeeds when the sector the adversary controls and
retains ends uncorrelated with the event---the fact is gone from what it holds, not
merely concentrated elsewhere in it; degrees of freedom discharged to the baths leave
the erasure target but remain in the entropy accounting. A third candidate, destroying
the correlation with nothing exported anywhere, is void by unitarity: distinguishable
global inputs remain distinguishable; what varies between strategies is where the
correlation goes and what entropy the operation produces. Intervening \emph{during} the
writing is a different problem---steering a monitored source so that its record reads
as the null history's is a control task, bounded in control effort rather than in work
for a continuously monitored qubit~\cite{Bugu2026}; the adversaries here arrive after
the records exist. A single notion with a single lower bound on work would be the
natural first guess; the model makes the split mandatory (Sec.~\ref{sec:adversary}).

\item \textbf{A fact at level $\bm{(R,\delta;k,W)}$.} An event recorded to objectivity
$(R,\delta)$ such that no operation in $\mathcal{U}(k,W)$ can operationally un-know it.
\end{itemize}

The accounting mechanisms are supported by two bookkeeping rules. First, \emph{register
origin}: a record written into a pre-reset (pure) register is not cheaper---its
cost was met at the reset. Costs are therefore counted in full, with the Landauer cost
of every register reset attributed to the record wherever and whenever the reset
occurred. (The per-memory seed of this rule is the Holevo--Landauer
mechanism~\cite{DebarbaHuberFriis2024}: memory mixedness bounds the entropy production
of redundant copying from below; the rule extends the accounting across the whole
preparation chain.) Second, \emph{bath neutrality}: information copied into degrees of
freedom that thermalize is not a record past its thermalization time; every record
carries a persistence horizon. Units, once: inside the entropy identities---the
inequality~(\ref{eq:master}) and the sector identity~(\ref{eq:sector})---entropies and
mutual informations are in nats; displayed laws carrying an explicit
$k_\mathrm{B}T\ln 2$ (the creation rule, the misalignment optimum, the maintenance
power) take $H(\varepsilon)$ and $H_2$ in bits, with the $\ln 2$ performing the conversion.

\section{The minimal broadcast model}\label{sec:model}

The model is a driven broadcast. $S$ is a qubit holding $\varepsilon$; the registers
$E_1,\ldots,E_N$ are qubits, each initially thermal at temperature $T$ with level
splitting $\Delta$, initially uncorrelated with $S$ and with each other. A work source
implements any sequence of controlled unitaries diagonal in $S$'s $z$-basis, with ideal
thermal contacts between strokes. The operation class is the model's fairness clause:
the einselecting interaction defines the machinery's basis, and forbidding bare
$S$-rotations is what ``recording machinery with interaction basis $z$'' means. All
ancillas and adversary memories enter the accounting on the same full-history terms as
the registers. Explicit costs below are quoted at the zero-gap benchmark $\Delta = 0$;
gapped registers enter through an exact two-part accounting (Sec.~\ref{sec:making})
whose total never falls below the bound. Within this class, every statement below is
exact; beyond it, the composed laws are developed, not established
(Sec.~\ref{sec:outlook}).

\section{Creation of a record}\label{sec:making}

For any protocol in the model class, the work obeys
\begin{equation}
  W \ge k_\mathrm{B}T \Big[ \sum_i I(\varepsilon{:}E_i) + I_{\mathrm{fin}}(\varepsilon{:}S)
  - H(\varepsilon) \Big],
  \label{eq:master}
\end{equation}
from the free-energy decomposition of the final state: thermal registers minimize local
free energy, so every local term is non-negative, and monotonicity bounds the total
correlation from below by the per-fragment information sum when the source is preserved.
When the system's own record is preserved, objectivity
at $(R,\delta)$ therefore requires
\begin{equation}
  W \ge R\, k_\mathrm{B}T \ln 2\, (1-\delta)\, H(\varepsilon):
  \label{eq:making}
\end{equation}
linear in redundancy at the Landauer coefficient, reduced by the deficit. Each demanded
record is a demanded quantity of correlation, and correlation against thermal
surroundings costs $k_\mathrm{B}T$ per nat~\cite{Huber2015}.

The bound is attained. For a uniform event, quasistatic reset-then-copy protocols
saturate Eq.~(\ref{eq:making}) in the slow limit, with finite-speed excess $c_1/M$ per
register for the $M$-step staircase,
$c_1 = k_\mathrm{B}T \ln[(1-e)/e]\,(\tfrac{1}{2}-e)/2$ at record error $e$. (The
single-measurement attainability precedent is Latune--Elouard's reversible
protocol~\cite{LatuneElouard2025}; here we set the redundancy.) The deficit reduction is
exactly of rate--distortion form under the same uniform restriction: in the
symmetric-error convention the per-register work reads
$k_\mathrm{B}T\ln 2\,[H(\varepsilon) - H_2(e)]$, and for uniform events the two
conventions coincide under $\delta = H_2(e)$. The attainability states are commuting
classical--quantum records, orthogonal---hence spectrum-broadcast---only in the perfect
limit.

For a biased event the lower bound stands unchanged but saturation at $k_\mathrm{B}T$
per nat fails, and the failure is physical: a perfect record requires orthogonal
conditional register states, whose local free energy raises the work of a perfect
record to a full $k_\mathrm{B}T\ln 2$ regardless of $H(\varepsilon)$. The register must
carry the event's whole distinguishing capacity, not its surprise. This is the work
form of the information-exchange accounting evaluated in the
model~\cite{SagawaUeda2012,EspositoVandenBroeck2011,ReebWolf2014}; the third-law
obstruction to pure pointers~\cite{Guryanova2020,DebarbaHuberFriis2024} sits adjacent to
it. The statement concerns the work of creation, not dissipation: with source access
the whole $k_\mathrm{B}T\ln 2$ is recoverable by uncomputation, and without it only the
margin $k_\mathrm{B}T\ln 2\,[1 - H(\varepsilon)]$ is---%
$k_\mathrm{B}T\ln 2\, H(\varepsilon)$ is the irreducible cost of a source-blind cycle.

The composition relies on two structural facts. Thermal registers minimize local free
energy, so no protocol extracts anything from the environment's initial state; and
total correlation dominates the per-fragment sum for \emph{any} fragment structure, so
collective encodings cannot do better---linearity is strict, by the same two-line
argument that proves Eq.~(\ref{eq:master}). Registers with a gap $\Delta$ carry an
exact two-part accounting: their initial thermal polarization reduces the writing work,
and holding them away from equilibrium afterward adds at least that much back, so the
full-history total never falls below the bound---cold or gapped media relocate the cost
of a record; they do not reduce it. If the protocol may consume the system's own
record, the bound gains a $-H(\varepsilon)$ term, the event's own record counted once.

The relation to the Holevo--Landauer bound~\cite{DebarbaHuberFriis2024} (also asserted
in bare counting form~\cite{Tank2025}) is exact: the per-register work inequality asserted here and
the entropy-production bound in~\cite{DebarbaHuberFriis2024} are two faces of one correlation/free-energy
decomposition, read in two different accountings. Their $\Sigma$ takes the memory as
the thermal receiver, in the system--reservoir decomposition of Esposito, Lindenberg,
and Van den Broeck~\cite{EspositoLindenbergVandenBroeck2010}; the quasistatic protocol in this work
creates positive Holevo information at zero \emph{total} entropy production, and the
distinction between the two entropy productions is what makes both statements true at
once.

\section{Recording off the pointer basis: the optimal work and a critical angle}\label{sec:basis}

Consider a uniform event in a basis rotated by $\theta$ from the machinery's basis. Every allowed operation commutes
with $z$-dephasing on $S$, so the records of all fragments jointly obey the ceiling
$I \le 1 - H_2(q)$ bits with $q = \sin^2(\theta/2)$: objectivity at deficit $\delta$ is
\emph{impossible at any work} beyond the critical angle
\begin{equation}
  H_2\!\big(\sin^2(\theta_c/2)\big) = \delta
  \quad\Longleftrightarrow\quad
  \theta_c = 2\arcsin\!\sqrt{H_2^{-1}(\delta)},
  \label{eq:border}
\end{equation}
with $\theta \in [0,\pi/2]$ and $H_2^{-1}$ the principal branch on $[0,1/2]$. The
ceiling and the critical angle are statements of quantum Darwinism: that only pointer
observables leave redundant imprints, with a finite-deficit criterion under which a
rotated observable has redundancy one, is Ollivier, Poulin, and Zurek's
theorem~\cite{OllivierPoulinZurek2005} (their Eqs.~(36)--(38)), resting on the
accessible-information complementarity of Zwolak and Zurek~\cite{ZwolakZurek2013}; the
dephasing-covariance route re-proves the criterion inside the model and yields the
critical angle as its explicit corollary.

\begin{figure}
\includegraphics[width=\columnwidth]{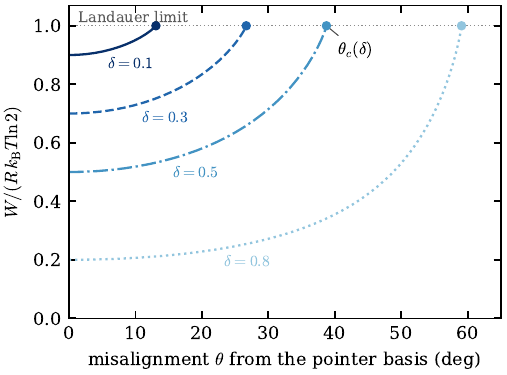}
\caption{Optimal work for recording off the pointer basis, Eqs.~(\ref{eq:border})
and~(\ref{eq:tariff}): the work per record to store a uniform event in a basis
misaligned by $\theta$ with the einselecting interaction, at four deficits. Each curve
rises from its rate--distortion value $1-\delta$ at $\theta = 0$ to the Landauer limit
exactly at its critical angle $\theta_c(\delta)$ (dots); beyond $\theta_c(\delta)$,
objectivity at deficit $\delta$ is unreachable at any work.}
\label{fig:tariff}
\end{figure}

Below $\theta_c$, what is new is the cost. The optimal work is
\begin{equation}
  \begin{gathered}
  W(\theta,\delta,R) = R\, k_\mathrm{B}T \ln 2\, \big[1 - H_2(e_1^*)\big], \\[2pt]
  e_1^* = \frac{e^* - q}{1 - 2q}, \qquad e^* = H_2^{-1}(\delta),
  \end{gathered}
  \label{eq:tariff}
\end{equation}
attained by sample-then-copy at the computed error split, and exactly optimal in the
model's operation class by a three-step argument: Mrs.~Gerber's
Lemma~\cite{WynerZiv1973} forces every fragment to carry at least $1 - H_2(e_1^*)$ bits
of correlation with the dephased bit; the work of any protocol is identical on the real
and dephased inputs, because register energetics are $z$-block functionals and the
$S$-side Hamiltonian is trivial; and the operation class preserves the dephased bit, so
the dephased-world bound applies to that correlation with no slack. The optimal work
rises from the rate--distortion value at $\theta = 0$ with leading behavior
$\propto \sin^2(\theta/2)$ and reaches $k_\mathrm{B}T\ln 2$ exactly at $\theta_c$
(Fig.~\ref{fig:tariff}): as misalignment consumes the available deficit, records must
approach perfection, and past $\theta_c$ perfection does not suffice. \emph{The pointer
basis is the argmin of the work of objectivity}---the work counterpart of the
predictability sieve, which selects pointer states by minimal decoherence-generated
entropy~\cite{ZurekHabibPaz1993,Zurek1993}. Work creates records of the pointer
observable at the fixed rates above~\cite{Zurek2022}; beyond $\theta_c$, no amount of
work creates records of a misaligned one.

The back-action completes the complementarity. For the minimum-disturbance family of
independent partial copiers, recording $\varepsilon$ in $z$ at redundancy $R$
multiplies $S$'s conjugate coherence by $\cos^R\!\varphi$, where $\cos\varphi$ is the
overlap of the conditional register states each copy leaves---the overlap decoherence
factor of Ref.~\cite{Zurek1982}. Conjugate coherence decays exponentially in
redundancy, and the factor depends on how the records were created, never on where they
are sent. Destination-independence is no-signalling, and the drift filter of
Sec.~\ref{sec:instruments} exhibits it in hardware. (Pointer selection when the event's
own dynamics does not commute with the monitoring is the neighbouring problem, treated
by Chisholm, Palma, and Innocenti~\cite{ChisholmPalmaInnocenti2026}.)

\section{Un-making: access, consolidation, and extraction}\label{sec:adversary}

One may expect that un-knowing obeys a lower bound on work of its own---the
Landauer cost of decorrelating every surviving record. We find this is not the case.

\textbf{The access theorem.} Reducing objectivity below $(R_{\min},\delta')$ requires
access to at least $k_{\min} = R - R_{\min} + 1$ of the fragments whose initial records
qualify at the target $\delta'$; with fewer, it is impossible at any work. Making a fact
needs $R$ successes anywhere; unmaking needs completeness everywhere. Objectivity is a
threshold property before it is an energetic one.

\textbf{Work protects nothing.} Given access, redundancy reduction can be
\emph{silent}: consolidating two perfect records by a CNOT ($b,b \to b,0$) lowers the
count unitarily at zero work and zero entropy production, and the fact survives,
concentrated, in the remaining records. The operation is Bennett-style reversible
compression~\cite{Bennett1982}, here in the adversary's hands. Full erasure of the
accessed records requires no work at all: XOR with a thermal ancilla and discard
exports the correlation into a bath-buried parity at $W = 0$ and $\Sigma = \ln 2$ per
record, so scrambled. And beyond the first copy, erasure yields net work: the $k-1$
inter-record parities of $k$ accessed perfect copies of a uniform bit are pure
registers, each worth $k_\mathrm{B}T\ln 2$ of extractable work, so an optimal adversary
extracts
\begin{equation*}
  W_{\mathrm{ext}} = (k-1)\, k_\mathrm{B}T\ln 2
\end{equation*}
while erasing all $k$ records, at total entropy production $\Sigma = \ln 2$---the
record-level instance of work locked in
correlations~\cite{PerarnauLlobet2015,Bera2017}. In the side-information accounting of
del Rio et~al.~\cite{delRio2011}, $k$ perfect copies of a uniform bit have joint entropy
$\ln 2$---one bit in $k$ carriers---so the two endpoints are consistent: resetting all
$k$ to blanks costs a single Landauer unit, while discharging them to full mixture
extracts $(k-1)\,k_\mathrm{B}T\ln 2$. Deep wells and gaps protect records against re-thermalization
(Sec.~\ref{sec:upkeep}), not against adversaries---given access, the only obstruction
is the access itself.

\textbf{What the second law constrains.} For the sector $A$ an adversary controls and
retains, with the source untouched and baths initially independent, the event-inclusive
entropy production obeys
\begin{equation}
  \Sigma_{\varepsilon A} = \Sigma_A
  + \big[ I_{\mathrm{i}}(\varepsilon{:}A) - I_{\mathrm{f}}(\varepsilon{:}A) \big]:
  \label{eq:sector}
\end{equation}
the event-inclusive dissipation equals the ordinary entropy production plus the
event-correlation destroyed. The identity is the second law with information
exchange---Sagawa and Ueda's
$\langle\sigma\rangle \ge \langle\Delta I\rangle$~\cite{SagawaUeda2012,SagawaUeda2013}
with the event as the untouched partner (their inequality holds without the
full-support assumption their integral equality needs, so perfect records are covered);
its continuous-flow form is Horowitz--Esposito~\cite{HorowitzEsposito2014}, its quantum
bipartite form Ptaszy\'nski--Esposito~\cite{PtaszynskiEsposito2019}, its earliest
bipartite form Allahverdyan--Janzing--Mahler~\cite{AllahverdyanJanzingMahler2009}, and
its subsystem decomposition
Esposito--Lindenberg--Van den Broeck~\cite{EspositoLindenbergVandenBroeck2010}. The
record-level reading is our contribution: erasing a fact from the sector dissipates at
least the correlation erased---$\Sigma_{\varepsilon A} \ge H(\varepsilon)$ for perfect
records, attained by the extraction protocol above---while consolidation, which
destroys no correlation, dissipates nothing. That is why the silence of consolidation
is consistent rather than paradoxical, and why ``work protects nothing'' coexists with
a second law that concedes nothing: the constraint is on correlation, not on work. For
noisy or biased records the constants are route-dependent---independent scrambling of
$k$ perfect records produces $\Sigma = k\ln 2$, and at record error $e = 0.1$ with
$k = 2$ the optimal extraction yields $W_{\mathrm{ext}} = 0.222\,k_\mathrm{B}T$ at
$\Sigma = 0.514$---and the extraction's reversibility does not extend unchanged beyond
the uniform perfect case. The creation-side Holevo--Landauer
bound~\cite{DebarbaHuberFriis2024} ($\langle\Sigma\rangle \ge H(X)$ at
spectrum-broadcast structure) depends on the same correlation bookkeeping, though
creation and source-blind sector erasure use different entropy accountings and access
assumptions: the connection is structural, not a time-reversal equivalence. With source
access the creation is fully reversible: uncomputation returns the registers to thermal
and the work to its source.

\textbf{Propagation.} When records propagate, the reach an adversary needs in order to
act on them grows with the event's age---radiative records recede at $c$, diffusive
ones as $\sqrt{Dt}$---while the fragment-count threshold of the access theorem is
unchanged by the records' motion. An event's records pass beyond reach through
propagation, not through energetics: as the event ages, the access requirement of the
bounded-$(k,W)$ analysis grows while its work requirement stays fixed.
Section~\ref{sec:instruments} does the calculation for a real instrument.

A remark on the arrow of time. Nothing here adds to the second law. One might locate
the past's immutability in an energetic bound; the model relocates it. Operational
objectivity is protected by access geometry---records recede at the speed of their
medium, and the reach required to act on an event's records grows calculably with its
age---while what thermodynamics constrains is erasure: destroying the sector's
correlation with the event dissipates at least the correlation destroyed,
Eq.~(\ref{eq:sector}). Redundancy can be quietly thinned; the fact itself cannot be
removed from what an adversary controls and retains except at that dissipation---and,
by unitarity, cannot be removed from the global state at all, only moved beyond the
sector.

\section{Maintenance: a fact is a nonequilibrium steady state}\label{sec:upkeep}

Records written into media that re-thermalize decay. Against register re-thermalization
at rate $\gamma$ (the polarization relaxation rate, $d(1-2e)/dt = -\gamma(1-2e)$), a
record of a uniform event with initial error $e_0$ stays within deficit $\delta$ for
$t_{\mathrm{hor}} = (1/\gamma)\ln[(1-2e_0)/(1-2e^*)]$, $e^* = H_2^{-1}(\delta)$, and
holding objectivity at $(R,\delta)$ indefinitely requires at least the power
\begin{equation}
  \begin{gathered}
  P = R\, k_\mathrm{B}T \ln 2\;\gamma\, g(\delta), \\[2pt]
  g(\delta) = \tfrac{1}{2}\,(1-2e^*)\log_2\frac{1-e^*}{e^*}.
  \end{gathered}
  \label{eq:upkeep}
\end{equation}
The per-register coefficient is the known two-state maintenance bound---Horowitz, Zhou,
and England's minimum driving cost at zero
gap~\cite{HorowitzZhouEngland2017}---evaluated at the record-quality boundary, and the
structural inequality behind it is the information-flow second law in rate
form~\cite{HorowitzEsposito2014}, in the family of the sensor learning-rate
bounds~\cite{BaratoHartichSeifert2014,HartichBaratoSeifert2014,BrittainJonesOuldridge2017,HorowitzSandberg2014}.
The objectivity-maintenance reading is this paper's: the map from the record constraint
$(R,\delta)$ to that boundary, the $R$-linearity, and optimality over \emph{all}
maintaining strategies. The boundary-refresh protocol attains the bound; the lower
bound follows from convexity of the bath's conditional entropy-production rate in the
record's conditional entropy, by Jensen's inequality, and extends over collective
strategies because conditional total correlation decreases under independent local bath
maps. The metrological
twin is recent: maintaining a state more parameter-sensitive than thermal requires a
steady entropy-production rate, in a Gaussian bosonic
setting~\cite{YoshimuraHamazaki2026}---Fisher information about a parameter bounded
where Eq.~(\ref{eq:upkeep}) bounds mutual information about an event.

The coefficient $g$ diverges logarithmically as $\delta \to 0$---perfect records have
unbounded maintenance rate---and vanishes as the deficit is fully relaxed,
$\delta \to 1$ ($e^* \to \tfrac{1}{2}$). Archival media enter by the Arrhenius
substitution $\gamma = \gamma_0 e^{-E_b/k_\mathrm{B}T}$, making $E_b/k_\mathrm{B}T$ the
archival quality; conservation-law-protected quantities are the infinite-well limit.
This deserves an interpretive statement: \emph{the past is actively maintained}.

\section{The speed of emergence}\label{sec:speed}

The creation rule bounds the work of objectivity and Eq.~(\ref{eq:upkeep}) the power of
keeping it; neither bounds its speed. Nature does not record quasistatically---sunlight
writes a dust grain's position into scattered photons in microseconds---so the direct
form of the roadmap's question is a rate law: how fast can objectivity emerge, per unit
dissipation?

The answer begins with a structural fact. At the zero-gap benchmark a thermal register
is uniform, and the uniform distribution is invariant under every permutation of its
levels, controlled or not: unitaries alone---including the copy gates---create no
record. Per register, let $D$ be the branch contrast (the difference between the two
conditional populations) and $M$ be the common polarization; common dissipative
dynamics only \emph{contracts} $D$, since both branches flow toward the same
instantaneous equilibrium, while an $\varepsilon$-controlled flip exactly exchanges $M$
and $D$. The functional $|D| + |M|$ therefore starts at zero and grows only through
bath flux: every record is written through the bath, with the controlled gate merely
converting common polarization into contrast. The Landauer step is the bottleneck in
time, not only in work.

Quantitatively, the statement is a speed limit. For any protocol in the model class
(per-register baths; $S$ as control only; all permutations, controlled or not, acting
on single registers) that takes $N$ uncorrelated zero-gap thermal registers to
objectivity $(R,\delta)$ for a uniform event,
\begin{equation}
  \Sigma\, A \;\ge\; \tfrac{1}{2}\, R^2\, (1-2e^*)^2,
  \qquad e^* = H_2^{-1}(\delta),
  \label{eq:speed}
\end{equation}
where $\Sigma$ is the event-inclusive entropy production (nats) and $A$ the total
dynamical activity---the mean number of bath jumps the protocol uses. The proof is
three Cauchy--Schwarz steps on one elementary bound: per branch,
$\dot\sigma \ge 2J^2/\dot a$ integrates to $\Sigma^b A^b \ge 2(L^b)^2$ with $L^b$ the
bath path length---the classical speed limit of Shiraishi, Funo, and
Saito~\cite{ShiraishiFunoSaito2018} in the optimal-transport product form unified by
Van Vu and Saito~\cite{VanVuSaito2023}; our contribution is the conversion argument and the
multi-record composition. The conversion argument gives $L^0 + L^1 \ge 1-2e^*$ per
record; averaging branches and summing records assembles Eq.~(\ref{eq:speed}). Rapid
recording requires dissipation or bath activity; neither alone is bounded below, but
their product is. Averaged over the preparation window---from the thermal start to
objectivity at time $\tau$, with mean entropy-production rate $\bar\sigma = \Sigma/\tau$
and mean activity rate $\bar{\mathfrak{a}} = A/\tau$---
\begin{equation*}
  \frac{R}{\tau} \;\le\; \frac{\sqrt{2\,\bar\sigma\,\bar{\mathfrak{a}}}}{1-2e^*}:
\end{equation*}
the average rate at which classicality emerges is bounded by the geometric mean of
dissipation and activity, a bound of thermodynamic-uncertainty
shape~\cite{BaratoSeifert2015}. The bound is a window statement, not a pointwise one: a
controlled flip can convert previously pumped polarization into records in an instant,
and the window is what accounts for the pumping. It is also a statement about scheduled
drives, not feedback loops: control conditioned on observed bath jumps can reach the
Landauer bound in finite time~\cite{LiuVanVu2026}, and in this accounting such a
monitor's own records are counted by the same rules---the feedback extension of
Eq.~(\ref{eq:speed}) is open.

At a common deadline $\tau$ and a per-register
activity cap, writing the $R$ records in parallel carries the bound
$\Sigma \ge R(1-2e^*)^2/(2\langle a\rangle\tau)$, while writing them sequentially
carries an extra factor of $R$---under those constraints, broadcasting is not merely
what environments do; it is the least-dissipation route to rapid redundancy. The
quasistatic limit is a trade: $\Sigma \to 0$ forces $A \to \infty$ above the bounded
product---the reversible protocols of Sec.~\ref{sec:making} substitute bath activity
for dissipation. The deficit enters as $(1-2e^*)^2$, a quality dependence different
from the creation rule's $(1-\delta)$ and the maintenance coefficient $g(\delta)$: the
three laws weigh record quality each in its own way.

The bound is sharp. A constant-affinity drive---the gap steered so that the log-ratio
of forward to backward bath flux holds at $f$---gives $\Sigma A$ exactly
$(f/2)\coth(f/2)$ times the bound, which tends to one as $f \to 0$: the relation is
asymptotically attained. At finite parameters, optimized six-segment protocols land
within ${\sim}10\%$ of the bound at the tested deficits (fitted deficit exponent
$2.0$), while the staircase protocol of Sec.~\ref{sec:making} sits a factor
${\approx}4$ above it.

Fragment-fragment gates lie outside the stated class, and they genuinely help: an
event-independent permutation of three registers can redistribute one bath-written
record into three weaker ones below the bound of Eq.~(\ref{eq:speed}). The three are
correlated---they agree more than they inform, the intersubjectivity distinction of
Sec.~\ref{sec:definitions} in thermodynamic form---so Eq.~(\ref{eq:speed}) governs
\emph{independent} redundancy, and the collective speed limit is an open problem; the
creation rule's strict linearity is untouched by the same construction. Coherent
protocols are conjectured to obey the same scaling, with the dynamical activity
suitably generalized;
gapped media enter through the two-part accounting of Sec.~\ref{sec:making}: a cold or
gapped register arrives already polarized, and the full-history accounting attributes
that polarization to its preparation. The bound applies to natural recording processes
once their entropy production and activity are identified---sunlight scattering off a
dust grain~\cite{RiedelZurek2010} is the canonical case.

\section{Instruments as record-making machines}\label{sec:instruments}

The framework meets measurement where per-event information budgets have been computed.
The adjacent framework---detectors as autonomous thermal machines whose detection
quality is tied to entropy production---exists at the minimal-model
level~\cite{Schwarzhans2026}; this section's axis is different (the performance
quantity is objectivity created: redundancy, deficit, persistence per joule), and its
instances are real instruments with computed per-event information budgets, which no
platform in that literature yet carries.

\textbf{The calorimeter.} A total-absorption calorimeter converts an event's energy
into a maximally redundant record---phonons, light, charge---and its per-event
information budget is published~\cite{ChungCalorimetry2026}: the instrument's resolution is set
by how well its \emph{kept} record constrains what it loses, and the resolution carried
by each readout dimension---a longer gate, finer cells, more light---is measured
channel by channel. On that program's hadron showers the one-observable record reads
$2.1\%$ at $20\,\mathrm{GeV}$ while the full record reaches $1.29\%$
($\approx 5.5\%/\sqrt{E} \oplus 0.3\%$ over 20--$100\,\mathrm{GeV}$), with dual readout
the largest single share. In the present frame the dictionary reads: the calorimeter's
resolution is the fidelity term $\delta$ of its records; its segmentation and light
yield are the redundancy $R$; its noise floors are the cost of holding the record
against the bath until readout---Eq.~(\ref{eq:upkeep}) with the instrument's own
$\gamma$ and $E_b/k_\mathrm{B}T$. The dictionary becomes quantitative once an event
ensemble, a fragment partition, and a relaxation model are fixed for the instrument;
fixing them is part of the design proposal below. One note from the Darwinism
literature must be mentioned: records written through a shared intermediate
apparatus inherit \emph{correlated} apparatus noise, which redundancy does not
mitigate~\cite{TranTranHo2026}---the instrument counterpart of the bath-neutrality rule,
and a design constraint this framing makes explicit.

\textbf{The drift filter.} The adiabatic drift filter is the opposite limit---a
minimal record-making machine---and its per-event energetics are exactly
calculated~\cite{ChungQuantumFilter2026}. All numbers below are at the filter's reference
operating point: an $18.6\,\mathrm{keV}$ electron at $1\,\mathrm{T}$, field decay
length $\lambda = 7\,\mathrm{cm}$, slow-drive geometry $\cot(y_0/\lambda) = 1$,
coherent-scale entrance ($\Delta Y_0 = \ell_{B0}/\sqrt{2}$). Each transit unavoidably
radiates $E_{\mathrm{rad}} \approx 0.66\,\mathrm{meV}$, carried by ${\sim}9$ microwave
photons near the $28\,\mathrm{GHz}$ entry quantum; the radiated energy scales as
$1/\cot(y_0/\lambda)$ across drive geometries (${\approx}0.20\,\mathrm{meV}$ in
${\sim}2.6$ photons at the companion paper's test geometry), and is energy-independent
across balanced settings at non-relativistic order. The emission is irreducible, and
each photon is an escaping record of the passage. The record is grained at the local
cyclotron quantum, and the induced uncertainty on the electron's own energy,
$\sigma^2 = \tfrac{3}{4}\,\hbar\omega_{c0}\,E_{\mathrm{rad}}\,c_\gamma$ with
$c_\gamma \approx 0.97$ the relativistic-harmonic dressing constant, is the shot noise
of those grains---the back-action, on the recorded electron, of the records'
existence. The thermal environment corrupts the records with a quartic turn-on below
the entry-quantum temperature, $1.34\,\mathrm{K}$ per tesla---a sharply calculable
instance of the bath-neutrality rule. And the filter's mode-independence theorem---the
same noise reaches any readout---is, in this frame, the statement that the record, once
emitted, is objective: its back-action does not depend on who collects it, the
destination-independence of Sec.~\ref{sec:basis} in hardware.

\textbf{Record propagation, worked.} These photons are the clean example of the growth
laws, and the flight arithmetic is exact. Each emitted record leaves at $c$ while its
electron drifts at $v_z = 2.6\times10^5\,\mathrm{m/s}$: a record outruns its own
electron by a factor ${\sim}1.1\times10^3$, and an adversary of reach $r$ cannot act on
a record once it has flown for $r/c$---$33\,\mathrm{ns}$ at laboratory scale,
$r = 10\,\mathrm{m}$. The event's records escape on a slower clock than any single
flight, and the two must be distinguished: the transit emits over its own duration
(photon rate $\propto e^{-2t/\tau_\lambda}$, $\tau_\lambda = 268\,\mathrm{ns}$), so
under free-space escape the Poisson stream puts the first record beyond
$10\,\mathrm{m}$ with $99\%$ probability only after ${\sim}140\,\mathrm{ns}$, with the
full complement following on the $\tau_\lambda$ scale. Under free escape, then, the
transit's records pass beyond laboratory reach on the timescale of the transit itself,
not before it. The $(R,\delta)$ these ${\sim}9$ photons constitute, given the
field-mode partition and background, is an estimation-theory question posed, not
computed, here; in the real instrument the photons terminate in the readout within tens
of ns, so free escape is an assumption, and the in-apparatus continuation (absorption,
diffusive spreading as $\sqrt{Dt}$, the enclosure's thermal re-emission) is proposed,
not computed. The interception surface grows as $4\pi c^2 t^2$; all timing content is
envelope-level, fine gyrophase timing being obstructed at the measured second
harmonic~\cite{ChungQuantumFilter2026}; the entry-quantum temperature sets the thermal
background for records in flight, and archival lifetime on arrival is a separate
material question, not claimed here.

Between the two machines sits this section's proposal: \emph{the thermodynamic budget
of objectivity as an instrument design quantity}. A detector is specified today by
resolution; in this frame it is specified by the $(R,\delta,\text{persistence})$ of the
records it creates per joule, with the laws above as the conversion---Eq.~(\ref{eq:making})
the minimum work, $\gamma\, g(\delta)$ the maintenance, and the propagation geometry
the time during which the records remain within reach.

\section{Outlook}\label{sec:outlook}

Is there a thermodynamic price associated with the emergence of classicality? In the
model the answer has three parts. Creating objectivity requires work---%
Eq.~(\ref{eq:making}), linear in redundancy, with strict linearity, rising with
misalignment to impossibility at $\theta_c$---but that work is stored rather than
dissipated: transferred quasistatically into thermal registers, it remains recoverable
with source access, and the creation can proceed at zero total entropy production in
the reversible limit. What cannot be made zero is everything else: speed
(Eq.~(\ref{eq:speed})---dissipation or bath activity, their product bounded below),
operation without the source ($k_\mathrm{B}T\ln 2\, H(\varepsilon)$ per cycle), and
maintenance---Eq.~(\ref{eq:upkeep})---for as long as the records persist. Emergence can
be made reversible; persistence cannot: a classical world that remains classical
dissipates continuously. And un-making is not creation reversed: records can be thinned
only through access, and removed from a sector only with dissipation at least the
correlation destroyed, Eq.~(\ref{eq:sector}).

Beyond the model class these laws are developed, not established, and four developments
are the natural next steps: an autonomous record-making machine---two baths, no
external drive, records as output---the counterpart for facts of the autonomous clocks;
a record-network formalism in which fragments and channels form a graph and the access
threshold becomes a min-cut statement; the thermodynamic closing of the
observer-reversal window in Wigner's-friend scenarios, whose dynamical precedent in a
Darwinism setting exists~\cite{RivlinEngineerBaumann2025}; and the estimation-theoretic
treatment of the drift filter's radiation, which would promote its ${\sim}9$ photons
from a count to a computed $(R,\delta)$.

We state three questions which can now be made precise and studied in future works. Where the
causal cones of records---light cones for radiative records, Lieb--Robinson cones in
lattices~\cite{LiebRobinson1972}---are cut by horizons, the un-making of infalling
facts and laboratory erasure acquire a common description: is the accounting continuous
across that comparison? Where locality is derived rather than assumed, is record
structure---which fragments \emph{can} carry which facts---doing the deriving? And the
vacuum: fluctuations that leave no redundant record are, in this frame, precisely
non-facts; is that emptiness of record structure a usable characterization?

Autonomous clocks gave the cost of recording time its own laws~\cite{Erker2017}; this
paper asks the same question of recording events, and its answers now stand where the
tick laws once stood---stated, attained in a minimal model, and waiting for a
measurement. The platforms in which redundant records have been
observed~\cite{Ciampini2018,Chen2019,Unden2019} are engineered environments with
controllable fragments, and the misalignment optimum, the extraction gain, and the
maintenance power are statements about work flows in exactly such systems. An event's
record, like a clock's tick, is a physical object with a computable cost. We have
computed the minimal costs for the simplest recorder; detectors already meet them, by
design; and the thermodynamic budget of objectivity belongs, we propose, on their
specification sheets.

\bibliography{refs}

\end{document}